\documentclass{article}
\usepackage{emulateapj}
\begin{document}

\submitted{To appear in \textit{The Astronomical Journal}}
\lefthead{PHOTOMETRICITY MONITOR AT APO}
\righthead{HOGG ET AL}

\title{A photometricity and extinction monitor at
       the Apache Point Observatory\altaffilmark{1}}
\author{       David W. Hogg\altaffilmark{2}}
\affil{        Department of Physics, New York University,
               4 Washington Place, New York, NY 10003\\
       \texttt{hogg@physics.nyu.edu}\\~}

\author{       Douglas P. Finkbeiner\altaffilmark{3,4},
               David J. Schlegel,
	       James E. Gunn}
\affil{        Department of Astrophysics, Princeton University,
               Princeton, NJ 08544\\
       \texttt{dfink, schlegel, jeg@astro.princeton.edu}\\~}

\altaffiltext{1}{Based on observations obtained at the Apache Point
                 Observatory, which is owned and operated by the
                 Astrophysical Research Consortium.}
\altaffiltext{2}{and Institute for Advanced Study, Princeton, NJ 08540}
\altaffiltext{3}{and Department of Astronomy, University of California
                 at Berkeley, 601 Campbell Hall, Berkeley, CA 94720}
\altaffiltext{4}{Hubble Fellow}

\begin{abstract}
An unsupervised software ``robot'' that automatically and robustly
reduces and analyzes CCD observations of photometric standard stars is
described.  The robot measures extinction coefficients and other
photometric parameters in real time and, more carefully, on the next
day.  It also reduces and analyzes data from an all-sky
$10~\mathrm{\mu m}$ camera to detect clouds; photometric data taken
during cloudy periods are automatically rejected.  The robot reports
its findings back to observers and data analysts via the World-Wide
Web.  It can be used to assess photometricity, and to build data on
site conditions.  The robot's automated and uniform site monitoring
represents a minimum standard for any observing site with queue
scheduling, a public data archive, or likely participation in any
future National Virtual Observatory.
\end{abstract}


\section{Introduction}

More and more, astronomical research is being performed remotely, in
the sense that the observer, or perhaps more properly ``data
analyst'', is now often not present at the place or time at which
observations are taken.  The increase in remoteness has several
causes.  One is that for many observatories, telecommunication is
easier than travel, especially if telescope allocations are of short
durations.  Another is that several new telescopes are using or plan
to use queue-scheduling (eg, contributions to \cite{boroson96}; and
\cite{boroson98}, \cite{garzon98}, \cite{tilanus00}, \cite{massey00}),
for which observer travel is essentially impossible.  Some new
ground-based telescopes are partially or completely robotic (eg,
contributions to \cite{filippenko92}, contributions to
\cite{adelman92}; and \cite{baruch93}, \cite{akerlof99},
\cite{castro-tirado99}, \cite{strassmeier00}, \cite{querci00}).
Possibly the most important reason for the increase in remote
observing is that many observatories, many large surveys, and some
independent organizations are creating huge public data archives which
allow analyses by anyone at any time (eg, contributions to
\cite{mehringer99} and \cite{manset00}).  There has been community
discussion of a ``National Virtual Observatory'' which might be a
superset of these archives and surveys (eg, contributions to
\cite{brunner01}).

Remote observing and archival data analysis bring huge economic and
scientific benefits to astronomy, but with the significant cost that
the observer does not have direct access to observing conditions at
the site.  Most remote observatories and data archives keep logs
written by telescope operators, but these logs are notoriously
non-uniform in their attention to detail and use of terminology.  All
sites that plan to host remote observers or maintain public data
archives must have repeatable, quantitative, astronomically relevant
site monitoring.

For these reasons, among others (involving photometric calibration and
direction of survey operations) the Sloan Digital Sky Survey (SDSS;
\cite{york00}), which is constructing a public database of
$10^4~\mathrm{deg}^2$ of five-bandpass optical imaging and $10^6$
optical spectra, employs several pieces of hardware for monitoring of
the Apache Point Observatory site, including a $10$-$\mathrm{\mu m}$
cloud-camera scanning the whole sky (\cite{hull94}), a single-star
atmospheric seeing monitor, a low-altitude dust particle counter, a
basic weather station, and a 0.5-m telescope making photometric
measurements of a large set of standard stars.  This paper is about a
fully automated software ``robot'' that reduces the raw 0.5-m
telescope data, locates and measures standard stars, and determines
atmospheric extinction in near-real time, reporting its findings back
to telescope and survey operators via the World-Wide
Web\footnote{Information about the WWW is available at
``http://www.w3.org/''.} (WWW).  This paper describes the software
robot, rather than the hardware, which will be the subject of a
separate paper (Uomoto et al in preparation).

Although the robot is somewhat specialized to work with the hardware
and data available at the Apache Point Observatory, it could be
generalized easily for different hardware.  We are presenting it here
because it might serve as a prototype for site monitors that ought to
be part of any functional remote observing site and of the National
Virtual Observatory.

\section{Telescope and detector}

The primary telescope used in this study is the Photometric Telescope
(PT) of the SDSS, located at the Apache Point Observatory (APO) in New
Mexico, at latitude 32\,46\,49.30\,N, longitude 105\,49\,13.50\,W, and
elevation 2788~m.  The PT has a 20-in primary mirror and is outfitted
with a $2048\times 2048$~pixel CCD with 1.16~arcsec pixels, making for
a $40\times 40~\mathrm{arcmin^2}$ field of view.  The telescope and
CCD will be described in more detail elsewhere (Uomoto et al in
preparation).

The telescope takes images through five bandpasses, $u'$, $g'$, $r'$,
$i'$, and $z'$, chosen to be close to those in the SDSS 2.5-m imaging
camera.  Filter wavelengths are given in Table~\ref{tab:bandpasses}.
The magnitude system here is based on an approximation to an AB system
(\cite{oke83}, \cite{fukugita96}), again because that was the choice
for the SDSS imaging.  The photometric system will be described in
more detail elsewhere (Smith et al in preparation).  Nothing about the
function of the robot is tied to this photometric system; any system
can be used provided that there is a well-calibrated network of
standard stars.

\section{Observing strategy}

Site monitoring and measurement of atmospheric extinction is only one
part of the function of the Photometric Telescope (PT).  The PT is
being simultaneously used to take data on a very large number of
``secondary patches'' that will provide calibration information for
the SDSS imaging.  For this reason, the PT spends only about one third
of its time monitoring standard stars.  Its site monitoring and
atmospheric extinction measuring functions could be improved
significantly, in signal-to-noise and time resolution, if the PT were
dedicated to these tasks.

The observing plan for the night is generated automatically by a field
``autopicker'' which chooses standard star fields on the basis of (a)
observability, (b) airmass coverage, (c) intrinsic stellar color
coverage, and (d) number of calibrated stars per standard field.
Because the observing is very regular, with $u'$, $g'$, $r'$, $i'$,
and $z'$ images taken (in that order) of each field, it has been
almost entirely automated.  The only significant variation in
observing from standard field to standard field is that different
fields are imaged for different exposure times to avoid saturation.

The raw data from the telescope is in the form of images in the
Flexible Image Transport System format (FITS; \cite{wells81}).  The
raw image headers contain the filter (bandpass) used, the exposure
time, the date and UT at which the exposure was taken, the approximate
pointing of the telescope in RA and Dec, and the type of exposure (eg,
bias, flat, standard-star field, or secondary patch).  The robot makes
use of much of this information, as described below.

\section{Obtaining and manipulating the data}
\label{sec:software}

In principle, the photometricity monitoring software could run on the
data-acquisition computer.  However, in the interest of limiting
stress on the real-time systems, the photometricity robot runs on a
separate machine, obtaining its data by periodic executions of the
UNIX \texttt{rsync} software\footnote{Information about \texttt{rsync}
is available at ``http://rsync.samba.org/''.}.  The \texttt{rsync}
software performs file transfer across a network, using a
remote-update protocol (based on file checksums) to ensure that it
only transfers updated or new files, without duplicating effort.  The
output of \texttt{rsync} can be set to include a list of the files
which were updated.  This output is passed (via scripts written in the
UNIX shell language \texttt{bash}\footnote{Information about
\texttt{bash} is available at the Free Software Foundation at
``http://www.fsf.org/''.}) to a set of software tools written in the
data-analysis language \texttt{IDL}\footnote{Information about
\texttt{IDL} is available at ``http://www.rsinc.com/''.}.

Virtually all of the image processing, data analysis, fitting, and
feedback to the observer is executed with \texttt{IDL} programs in
\texttt{bash} wrappers.  This combination of software has proven to be
very stable and robust over many months of continuous operation.  In
addition, data reduction code written in \texttt{IDL} is easy to
develop and read.  Our only significant reservation is that
\texttt{IDL} is a commercial product which is not open-source.  We
have responded to this lack of transparency by writing as much as
possible of the robot's function in terms of very low-level
\texttt{IDL} primitives, which could be re-written straightforwardly
in any other data-analysis language if we lost confidence in or lost
access to \texttt{IDL}.  We have not found the lack of transparency to
be limiting for any of the functionality described in this paper.

\section{Correcting the images}
\label{sec:correcting}

Each raw image is bias-subtracted and flattened, using biases and
flat-field information archived from the most recent photometric
night.  How the bias and flat images are computed from a night's data
is described in Section~\ref{sec:nextday}, along with the conditions
on which a night is declared photometric.

\begin{figure}
%
%
\plotone{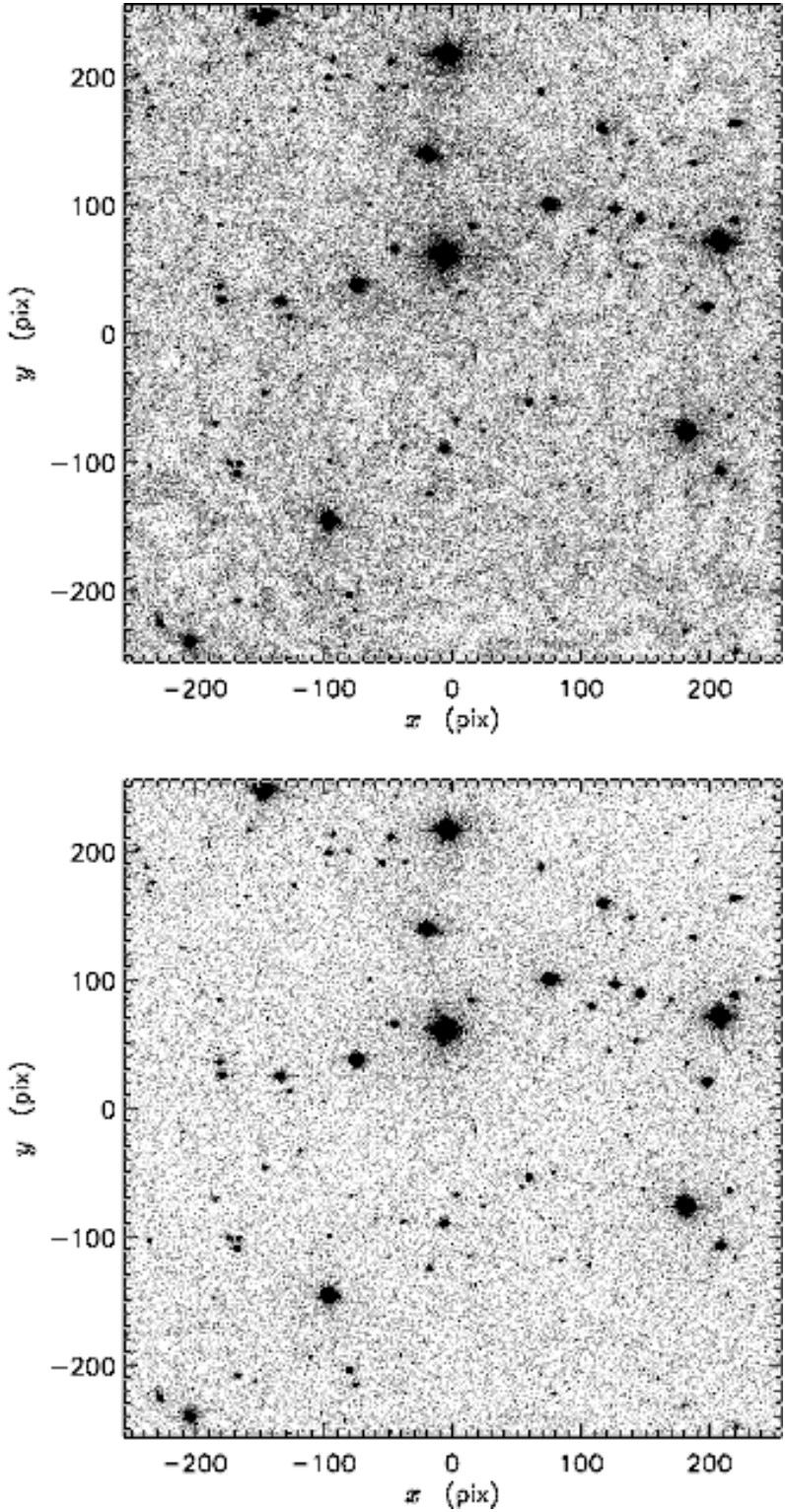}
\figcaption[Demonstration of fringing correction]{A $512\times
512~\mathrm{pix^2}$ ($594\times 594~\mathrm{arcsec^2}$) section of a
bias-subtracted and flat-fielded $z'$-bandpass image from the PT,
\textsl{(top)}~before, and \textsl{(bottom)}~after the fringing
correction.  The two images are stretched identically; in the
\textsl{(top)} image, the fringing is roughly 10~percent of the
background level, peak-to-peak.\label{fig:fringe_demo}}
\end{figure}

Because the CCD is thinned for sensitivity in the $u$ bandpass, it
exhibits interference fringing in the $i'$ and $z'$ bandpasses.  The
fringing pattern is very stable during a night and from night to
night.  The fringing is modeled as an additive distortion; it is
removed by subtracting a ``fringe image'' scaled by the DC sky level.
How the fringe image is computed is described in
Section~\ref{sec:nextday}.  The DC sky level is estimated by taking a
mean of all pixels in the image, but with iterated ``sigma-clipping''
in which the rms pixel value is computed, 3.5-sigma outlier pixels are
discarded, and the mean and rms are re-estimated.  This process is
iterated to convergence.  The fringing correction is demonstrated in
Figure~\ref{fig:fringe_demo}.  Although in principle the fringing
pattern may depend on the temperature or humidity of the night sky, it
does not appear to vary significantly within a night, or even
night-to-night.  Perhaps surprisingly, the fringing pattern appears
fixed and its amplitude scales linearly with the DC level of the sky.

For each corrected PT image, a mask of bad pixels is retained.  The
mask is the union of all pixels saturated in the raw images along with
all pixels which are marked as anomalous in the bias image, flat
image, or (for the $i'$ and $z'$ bandpasses) fringe image.  How pixels
are marked as anomalous in the bias, flat, and fringe is described in
Section~\ref{sec:nextday}.

The individual bias-subtracted, flattened and (for the $i'$ and $z'$
bandpasses) fringe-corrected image will, from here on, be referred to
as the ``corrected frames''.  The corrected frames are stored and
saved by the robot in FITS format.

\section{Astrometry and source identification}
\label{sec:astrom}

Hardware pointing precision is not adequate for individual source
identifications.  The software robot determines the precise astrometry
itself, by comparison with the USNO-SA2.0 astrometric catalog
(\cite{monet98}).  This catalog contains $5\times 10^7$ astrometric
stars over most of the sky; there are typically a few $10^2$ catalog
stars inside a PT image.

In brief, the astrometric solution for each image is found as follows:
A sub-catalog is created from all astrometric stars within 30~arcmin
of the field center (as reported by the hardware in the raw image
header).  An implementation of the DAOPHOT software (\cite{stetson87},
\cite{stetson92}) in \texttt{IDL} (by private communication from
W. Landsman) is used to locate all bright point sources in the
corrected frame (DAOPHOT is used for object-finding only, not
photometry).  Each set of stars is treated as a set of delta-functions
on the two-dimensional plane of the sky, and the cross-correlation
image is constructed.  If there is a statistically significant cluster
of points in the cross-correlation image, it is treated as the offset
between the two images.  Corresponding positions in the two sets of
stars (from the corrected frame and from the astrometric catalog) are
identified, and the offset, rotation, and non-linear radial distortion
in the corrected frame are all fit, with iterative improvement of the
correspondence between the two sets of stars.  The precise astrometric
information is stored in the FITS header of each corrected frame in
GSSS format (this format is not standard FITS but is used by the HST
Guide Star Survey; cf.\ \cite{russell90}, \cite{calabretta00}).  Our
algorithm is much faster, albeit less general, than previous
algorithms (eg, \cite{valdes95}).

Essentially all exposures of more than a few seconds in the $g'$, $r'$
and $i'$ bandpasses obtain correct astrometric solutions by this
procedure.  Some short $u'$ and $z'$ exposures do not, and are picked
up on a second pass using a mini-catalog constructed from stars
detected in the $g'$ and $r'$ exposures of the same field.  On most
nights, all exposures in all bands obtain correct astrometric
solutions.

The algorithm and its implementation will be discussed in more detail
in a separate paper, as it has applications in astronomy that go
beyond this project.

\section{Photometry}

The robot measures and performs photometric fits with the stars in the
SDSS catalog of photometric standards (Smith et al in preparation).
This catalog includes stars in the range $6<r<15$~mag, calibrated with
the USNO 40-inch Ritchey-Chr\'etien Telescope.  Several of the
standard-star fields used in this catalog are well-studied fields
(\cite{landolt92}) containing multiple standard stars spanning a range of
magnitude and color.

The photometric catalog is searched for photometric standard stars
inside the boundaries of each corrected frame with precise astrometric
header information.  If any stars are found in the photometric
catalog, aperture photometry is performed on the point source found at
the astrometric location of each photometric standard star.  The
centers of the photometric apertures are tweaked with a centroiding
procedure which allows for small ($\lesssim 1$~arcsec) inaccuracies in
absolute astrometry.  The aperture photometry is performed in
focal-plane apertures of radii 4.63, 7.43, and 11.42~arcsec.  The sky
value is measured by taking the mean of the pixel values in an annulus
of inner radius 28.20 and outer radius 44.21~arcsec, with iterated
outlier rejection at 3~sigma.  All of these angular radii are chosen
to match those used by the SDSS PHOTO software (\cite{lupton01},
Lupton et al in preparation).

In what follows, the 7.43-arcsec-radius aperture photometry is used.
This aperture was chosen from among the three for showing, on nights
of typical seeing, the lowest-scatter photometric solutions.  This is
because, in practice, the 7.43-arcsec-radius aperture is roughly the
correct trade-off between individual measurement signal-to-noise
(which favors small apertures) and insensitivity to spatial or
temporal variations in the point-spread function (which favors large).
The PT shows significant, repeatable, systematic distortions of the
point-spread function across the field of view; a more sophisticated
robot would model and correct these distortions; for our purposes it
is sufficient to simply choose the relatively large 7.43-arcsec-radius
aperture.

Each photometric measurement is corrected for its location in the
field of view of the PT.  There are two corrections.  The first is an
illumination correction derived from the radial distortion of the
field as found in the precise astrometric solution
(Section~\ref{sec:astrom}).  The illumination correction is designed
to account for the fact that photometrically we are interested in
fluxes, but the flatfield is measured with the sky; ie, the flatfield
is made to correct pixels to a constant surface-brightness sensitivity
rather than a constant flux sensitivity.  Because of optical
distortions, pixels near the edge of the CCD see a different solid
angle than pixels near the center.  Empirically, the dominant field
distortion appears radial, so no attempt has been made to correct for
illumination variation from arbitrary distortions.  The illumination
correction reaches a maximum of $\sim 0.02$~mag at the field edges and
$\sim 0.04$~mag at the field corners.

The second correction is related to the fringing in the $i'$ and $z'$
bandpasses.  The CCD is thinned, but not precisely uniformly; the
dominant thinning gradients are radial.  Because of reflections
internal to the CCD, gradients in CCD thickness lead to gradients in
the fraction of point-source light scattered out of the seeing core
and into the sky level.  Since the flat-field is computed on the basis
of sky level, these gradients are seen as residuals in photometry.
Radial photometry corrections for the $i'$ and $z'$ bandpasses were
found by performing empirical fits to photometry residuals; they are
applied to the $i'$ and $z'$ bandpass photometry.  These ``thinning''
corrections reach a maximum of $[\Delta i',\Delta z']\sim
[0.02,0.05]$~mag at the field edges and $\sim [0.03,0.07]$~mag at the
field corners.

\section{Cloud detection and data veto}
\label{sec:cloudveto}

\begin{figure}
%
%
\plotone{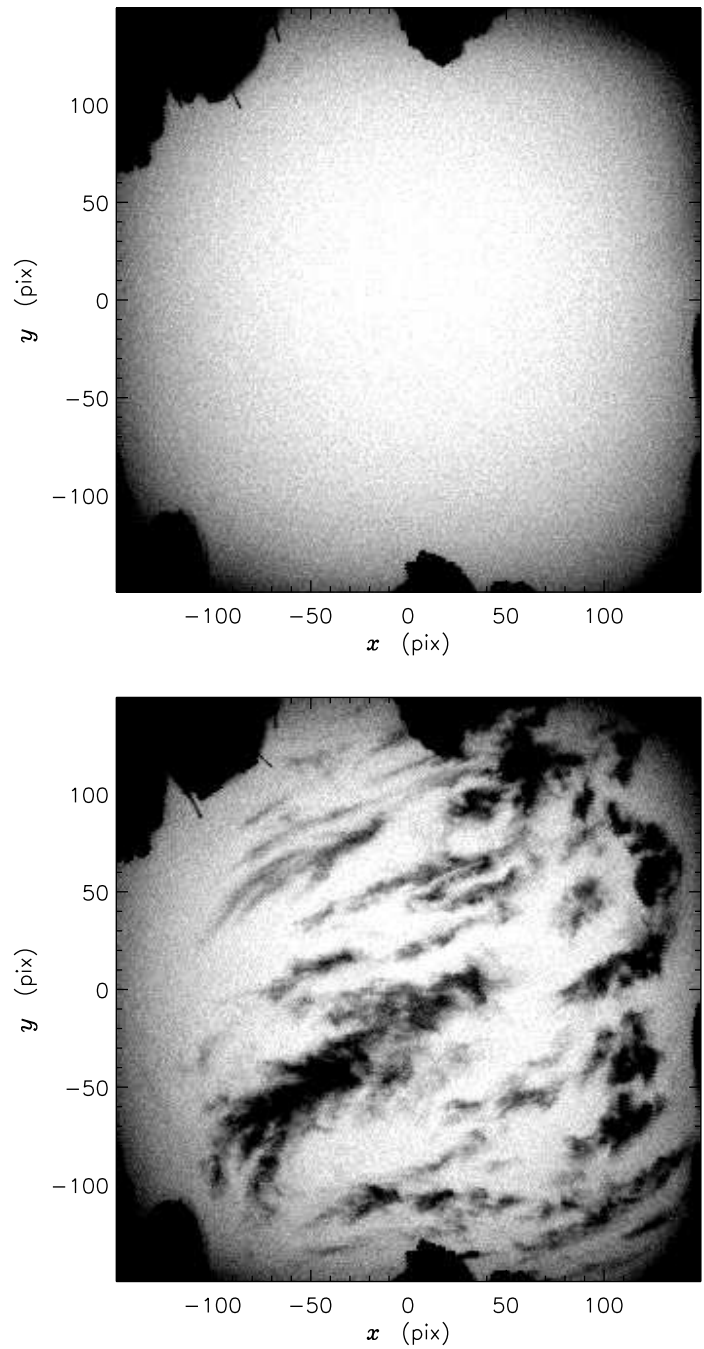}
\figcaption[Example cloud-camera images]{Two example cloud-camera
images, taken \textsl{(top)} at 05:24 UT on MJD 51997, and
\textsl{(bottom)} at 05:45 UT on MJD 51997.  The images are displayed
at the same stretch and are negatives (bright areas appear dark).  The
dark silhouettes around the edge are buildings and trees near the
camera; the APO 3.5-m Telescope is visible in the upper left-hand
corner.\label{fig:cloudcam}}
\end{figure}

The ``prototype cloud camera'' (\cite{hull94}) operating at APO
utilizes a single-pixel cooled $10~\mathrm{\mu m}$ detector and
scanning mirrors.  The sky is scanned by two flat mirrors driven by
stepper motors, followed by an off-axis hyperbolic mirror that images
the sky onto a single channel HgCdTe photoconductive
detector.\footnote{For more information on the cloud camera, see
http://www.apo.nmsu.edu/Telescopes/SDSS/sdss.html} The detector
samples 300 times per scan, 300 scans per image, yielding an image
with $300\times 300$~pixels covering a $135\times 135~\mathrm{deg^2}$
field with the $0.9$~deg beam.  Some typical images are shown in
\figurename~\ref{fig:cloudcam}.  An image is completed in
approximately 5~min.  This is somewhat slow for real-time monitoring,
but perfectly adequate for our purposes.  This design was prefered
over a solid state array for reasons of price, stability, and field of
view.  A disadvantage is the maintenance required by moving parts, but
this has not been a serious drawback.

Experimentation showed that a simple and adequate method for detecting
cloud cover is to compute the rms value of the sky within 45~deg of
zenith in each frame.  This seems to be a simple and robust method,
and it fails only in the case of unnaturally uniform cloud cover (this
has never occurred).  When the cloud-camera rms exceeds a predefined
threshold for a period of time, that period is declared bad, and the
data taken during that interval are ignored for photometric parameter
fitting.  The bad interval is padded by 20~min on each end, so that
even a single cloud appearing in one frame requires discarding at
least 40~min of data.  This is conservative, but it is more robust to
set the cloud threshold high and reject significant time intervals
than to make the threshold extremely low.

\section{Photometric solution}

Every time the PT completes a set of $u',g',r',i',z'$ exposures in a
field, the robot compiles all the measurements made of photometric
standard stars in that night and fits photometric parameters to all
data not declared bad by the cloud-camera veto
(Section~\ref{sec:cloudveto}).  The photometric equations used for the
five bandpasses are
\begin{eqnarray}
u'_\mathrm{inst} & = & u_\mathrm{USNO} + a_u + b_u\,(u-g)_\mathrm{USNO}
 + k_u\,X \nonumber \\
 & & + c_u\,[X-X_0]\,[(u-g)_\mathrm{USNO}-(u-g)_0] \nonumber \\
 & & + \dot{a}_u\,[t-t_0]] \nonumber \\
g'_\mathrm{inst} & = & g_\mathrm{USNO} + a_g + b_g\,(g-r)_\mathrm{USNO}
 + k_g\,X \nonumber \\
 & & + c_g\,[X-X_0]\,[(g-r)_\mathrm{USNO}-(g-r)_0] \nonumber \\
 & & + \dot{a}_g\,[t-t_0]] \nonumber \\
r'_\mathrm{inst} & = & r_\mathrm{USNO} + a_r + b_r\,(r-i)_\mathrm{USNO}
 + k_r\,X \nonumber \\
 & & + c_r\,[X-X_0]\,[(r-i)_\mathrm{USNO}-(r-i)_0] \nonumber \\
 & & + \dot{a}_r\,[t-t_0]] \nonumber \\
i'_\mathrm{inst} & = & i_\mathrm{USNO} + a_i + b_i\,(i-z)_\mathrm{USNO}
 + k_i\,X \nonumber \\
 & & + c_i\,[X-X_0]\,[(i-z)_\mathrm{USNO}-(i-z)_0] \nonumber \\
 & & + \dot{a}_i\,[t-t_0]] \nonumber \\
z'_\mathrm{inst} & = & z_\mathrm{USNO} + a_z + b_z\,(i-z)_\mathrm{USNO}
 + k_z\,X \nonumber \\
 & & + c_z\,[X-X_0]\,[(i-z)_\mathrm{USNO}-(i-z)_0] \nonumber \\
 & & + \dot{a}_z\,[t-t_0]]
\end{eqnarray}
where the $m_\mathrm{inst}$ symbolize instrumental magnitudes defined by
\begin{equation}
m_\mathrm{inst}\equiv -2.5\,\log_{10}\left(\frac{DN}{t_\mathrm{exp}}\right)
\end{equation}
with $DN$ the flux in raw counts in the corrected frame and
$t_\mathrm{exp}$ the exposure time; the $m_\mathrm{USNO}$ symbolize
the magnitudes in the photometric standard-star catalog; $X$
symbolizes airmass; $t$ symbolizes time (UT); $X_0$, $t_0$, and the
colors $(u-g)_0$, etc, symbolize fiducial airmass, time, and colors
(arbitrarily chosen but close to mean values); and the $a$, $b$, $k$,
$c$, and $\dot{a}$ parameters are, in principle, free to vary.  The
system sensitivities are the $a_i$; the tiny differences in
photometric systems between the USNO 40-inch and PT bandpasses are
captured by the color coefficients $b_i$; the atmospheric extinction
coefficients are the $k_i$; atmospheric extinction is a weak function
of intrinsic stellar color parameterized by the $c_i$; and the
$\dot{a}_i$ parameterize any small time evolution of the system during
the night.

The above photometric equations are not strictly correct for an AB
system, because in the AB system, there is no guarantee that the
colors of standard stars through two slightly different filter systems
will agree at zero color; this agreement is assumed by the above
equations.  In an empirical system, such as the Vega-relative
magnitude system, a certain star, such as Vega (and stars like it)
have zero color in all colors of all filter systems.  The AB system is
based on a hypothetical source with $f_{\nu}=\mathrm{const}$;
there is no guarantee that a source with zero color in one filter
system will have zero color in any other.  The offsets must be
computed theoretically, using models of CCD efficiencies, mirror
reflectivities, atmospheric absorption spectra, and intrinsic stellar
spectral energy distributions.  We have ignored this (subtle) point,
since it only leads to offsets in the sensitivity parameters $a_i$ and
does not affect photometricity or atmospheric extinction assessments.

In practice, the $c$ parameters are always fixed at theoretically
derived values
\begin{eqnarray}
c_u & = & -0.021~\mathrm{mag\,mag^{-1}\,airmass^{-1}} \nonumber\\
c_g & = & -0.016 \nonumber\\
c_r & = & -0.004 \nonumber\\
c_i & = & +0.006 \nonumber\\
c_z & = & +0.003
\end{eqnarray}
The $\dot{a}$ parameters are only allowed to be non-zero in the
next-day analysis (Section~\ref{sec:nextday}).  Because the design
specification on the PT did not require correct airmass values in raw
image headers, the airmass values are computed on the fly from the RA,
Dec, UT, and location of the Observatory (eg, \cite{smart77}).  The
reference airmass and colors are chosen to be roughly the mean on a
typical night of observing, or
\begin{equation}
X_0= 1.30~\mathrm{airmass}
\end{equation}
\begin{eqnarray}
(u-g)_0 & = & +1.42~\mathrm{mag} \nonumber\\
(g-r)_0 & = & +1.11 \nonumber\\
(r-i)_0 & = & +0.48 \nonumber\\
(i-z)_0 & = & +0.35
\end{eqnarray}
(in the AB system).  In the next-day analysis, the reference time
$t_0$ is set to be the mean UT of the observations.

In principle the color coefficients $b_i$ could be determined globally
and fixed once and for all.  However, the PT filters have been shown
to have some variation with time and with humidity (they are kept in a
low-humidity environment with some variability), so the robot fits
them independently every night.  Freedom of the $b_i$ is allowed not
because it is demanded by the data, but rather because measurements of
the $b_i$ over time are an important part of data and telescope
quality monitoring.  Typical color coefficient values (which measure
differences between the USNO 40-inch telescope and PT bandpasses) are
\begin{eqnarray}
b_u & = & +0.048~\mathrm{mag\,mag^{-1}} \nonumber\\
b_g & = & -0.004 \nonumber\\
b_r & = & -0.028 \nonumber\\
b_i & = & -0.053 \nonumber\\
b_z & = & -0.031
\end{eqnarray}

Quite a bit of experimentation went into the photometric equations.
Inclusion of theoretically determined $c$ parameters minutely improves
the fit, but on a typical night there are not enough data to determine
the $c$ parameters empirically; the $c_i$ are held fixed.  Terms
proportional to products of time and airmass, again, are not well
constrained from a single night's data; they are not included in the
fit.  To each bandpass a color is assigned; the choices have been made
to use the most well-behaved ``adjacent'' colors.  (The $(u-g)$ color
is not used for the $g$ equation because the boundary between $u$ and
$g$ is, by design, on the 4000~\AA\ break, making color
transformations extremely sensitive to the precise properties of the
filter curves.)

In the real-time analyses, the $a$, $b$, and $k$ parameters (15 total)
are fit to the counts from the aperture photometry (and USNO
magnitudes, times, airmasses, etc) with a linear least-square fitting
routine, which iteratively removes 3.0-sigma outliers and repeats its
fit.  A typical night at APO shows extinction coefficients
\begin{eqnarray}
k_u & = & 0.52~\mathrm{mag\,airmass^{-1}} \nonumber\\
k_g & = & 0.19 \nonumber\\
k_r & = & 0.12 \nonumber\\
k_i & = & 0.08 \nonumber\\
k_z & = & 0.06
\end{eqnarray}
with factor of $\sim 1.5$ variations from night to night.

All photometric measurements are weighted equally in these fits,
because the error budget appears to be dominated by systematic errors
in the USNO catalog and sky subtraction; the primary errors are not
from photon statistics.  As observing continues and iterative
improvement to the USNO catalog is made, we will be able to use a more
sophisticated weighting model; we don't expect such improvements to
significantly change the values of our best-fit photometric
parameters.

On a typical night, 5-band measurements are made of 20 to 50 standard
stars in 10 to 15 standard-star fields.

\section{Real-time observer feedback}
\label{sec:feedback}

\begin{figure}
\plotone{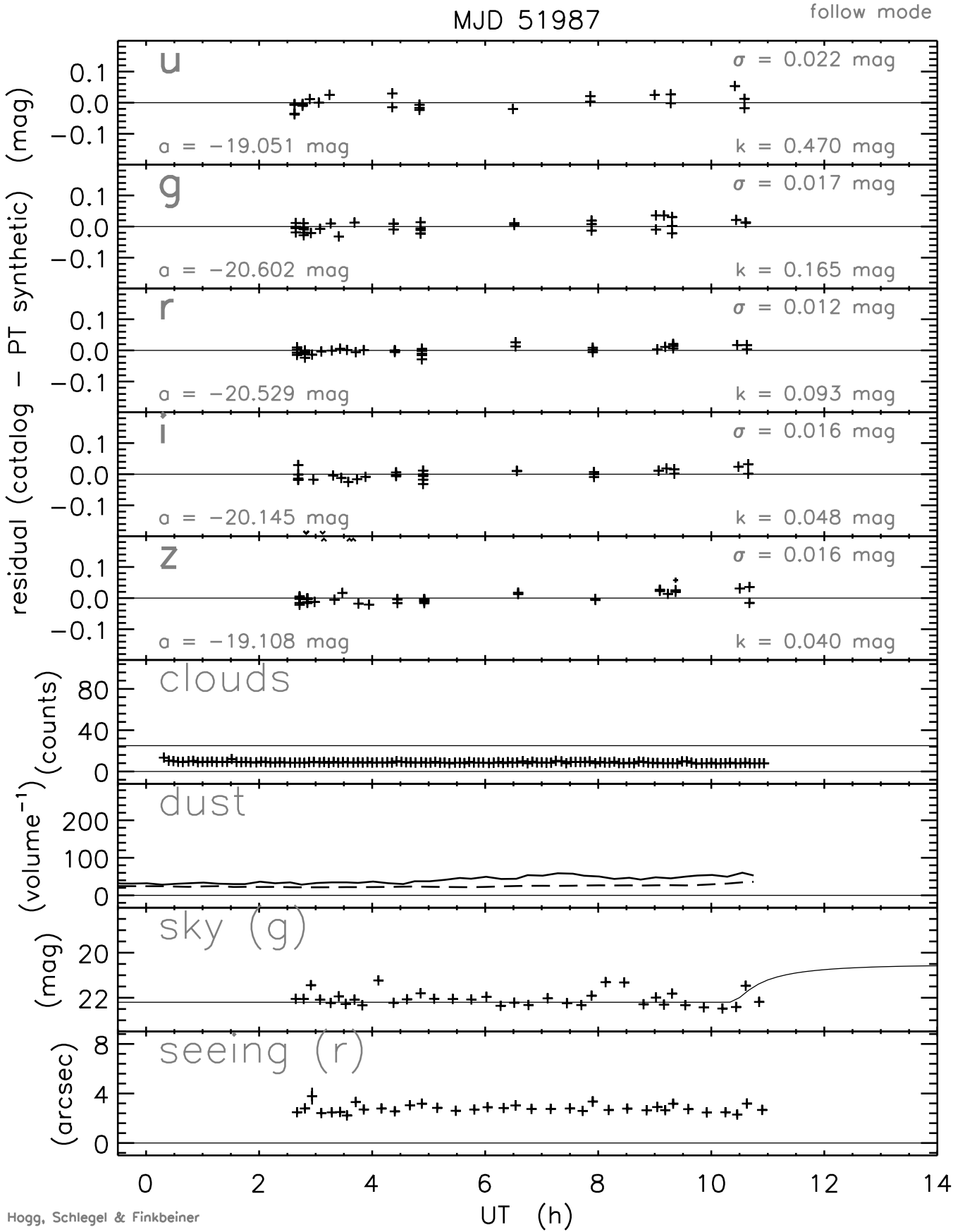}
\figcaption[Example of real-time WWW output]{The real-time WWW output
created at 11:00 UT on MJD 51987.  This was one of the best nights of
the year at APO.\label{fig:feedback1}}
\end{figure}

\begin{figure}
\plotone{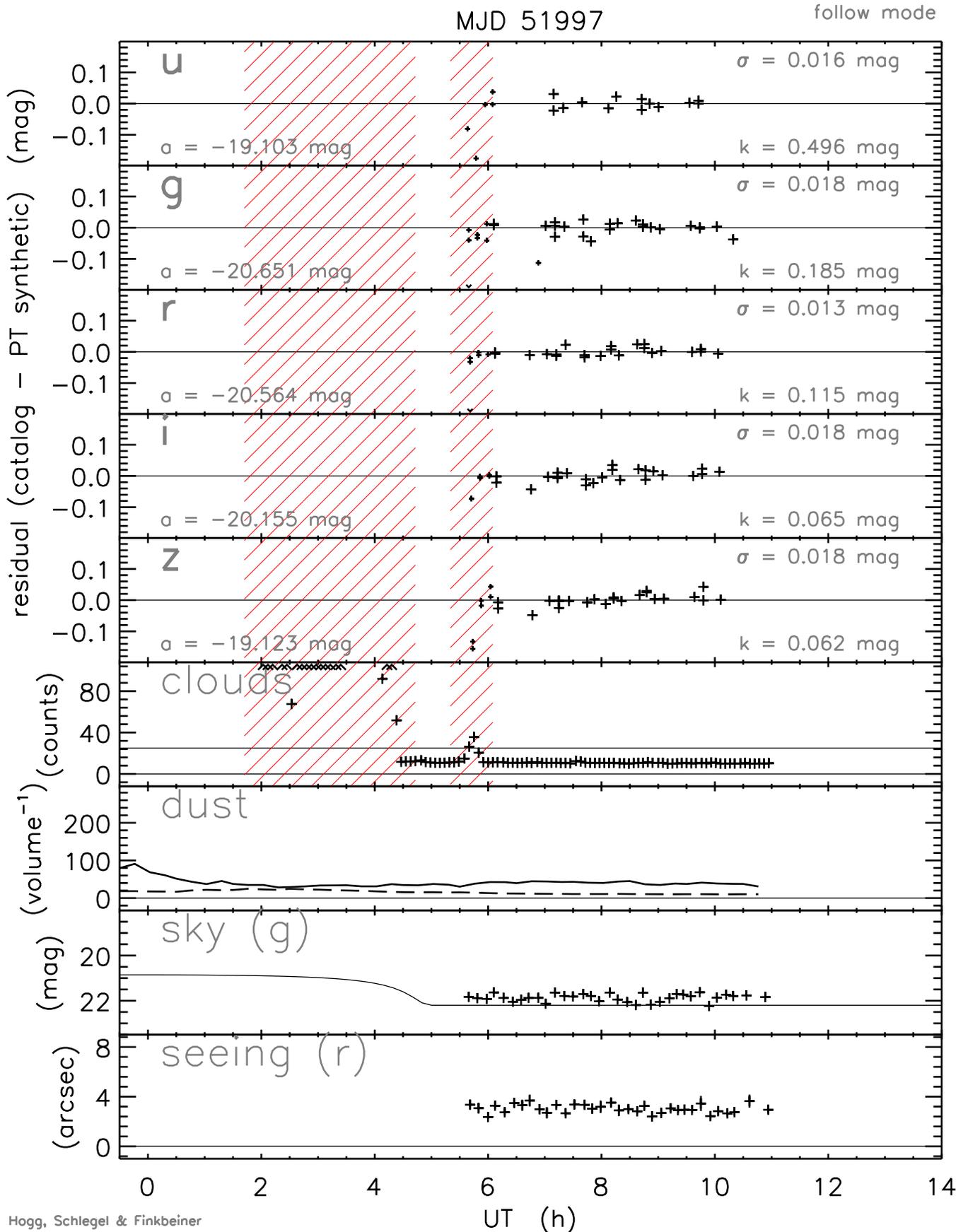}
\figcaption[Example of real-time WWW output]{The real-time WWW output
created at 11:00 UT on MJD 51997.  The red hatched areas are time
intervals vetoed by the cloud-camera analysis.  Rejected points are
shown with smaller symbols.\label{fig:feedback2}}
\end{figure}

Every ten minutes during the night (from 18:00 to 07:00, local time),
the robot builds a WWW page in Hypertext Markup
Language\footnote{Information about HTML is available at
``http://www.w3.org/''.} (HTML) reporting on its most up-to-date
photometric solution.  The WWW page shows the best-fit $a$, $b$ and
$k$ parameter values, plots of residuals around the best photometric
solution in the five bandpasses, and root-mean-squares (rms) of the
residuals in the five bandpasses.  Parameters or rms values grossly
out-of-spec are flagged in red.  This feedback allows the observers to
make real-time decisions about photometricity, and confirm
expectations based on visual impressions of and 10-$\mathrm{\mu m}$
camera data on atmospheric conditions.  The WWW page also includes an
excerpt from the observers' manually entered log file.  This allows
those monitoring the site remotely to compare the observers' and
robot's photometricity judgements.  Figures~\ref{fig:feedback1} and
\ref{fig:feedback2} show examples of the photometricity output on the
WWW page on two typical nights.

The WWW page also shows data from the APO weather, dust, and cloud
monitors for comparison by the observers, as do
Figures~\ref{fig:feedback1} and \ref{fig:feedback2}.

Since WWW pages in HTML are simply text files, they are trivially
output by \texttt{IDL}.  The figures on the WWW pages are output by
\texttt{IDL} as PostScript files and then converted to a (crudely)
antialiased pixel image with the UNIX command
\texttt{ghostscript}\footnote{Information about \texttt{ghostscript}
is available from ``http://www.gnu.org/''.} and the UNIX
\texttt{pbmplus} package\footnote{Information about the
\texttt{pbmplus} package is available from ``http://www.acme.com/''.}.

\section{Next-day analysis and feedback}
\label{sec:nextday}

At 07:00 (local time) each day, the robot begins its ``next-day''
analysis, in which all the data from the entire night (just ended) are
re-reduced and final decisions are made about photometricity and data
acceptability.

Bias frames taken during the night are identified by header keywords.
The bias frames are averaged, with iterated outlier rejection, to make
a bias image.  Any pixels with values more than 10-sigma deviant from
the mean pixel value in the bias image are flagged as bad in a bias
mask image.

Dome and sky flat frames are identified by header keywords.  These raw
images are bias-subtracted using the bias image.  Each bias-subtracted
image is divided by its own mean, which is estimated with iterated
outlier rejection.  The bias-subtracted and mean-divided flat frames
are averaged together, again with iterated outlier rejection, to make
five flat images, one for each bandpass.  Any pixels with values more
than 10-sigma deviant from the mean are flagged as bad in flat mask
images.

The $i$ and $z$-bandpass images are affected by an interference
fringing pattern, which is modeled as an additive distortion.  Any $i$
or $z$-bandpass image taken during the night with exposure time
$t_\mathrm{exp}>100$~s is identified.  These long-exposure frames are
bias-subtracted and divided by the flat.  Each is divided by its own
mean, again estimated with iterated outlier rejection.  These
mean-divided frames are averaged together, again with iterated outlier
rejection to make two fringe correction images, in the $i'$ and $z'$
bandpasses.  Again, 10-sigma outlier pixels are flagged as bad in
fringe mask images.

Constant bias, flat and fringe images are used for the entire night;
there is no evidence with this system that there is time evolution in
any of these.

All the raw images from the night are bias-subtracted and flattened
with these new same-night bias and flat images.  The $i'$ and
$z'$-bandpass images are also fringe-corrected.  Thus a new set of
corrected frames is constructed for that night.  The real-time
astrometry solutions are re-used in these new corrected frames.  All
photometry is re-measured in the new corrected frames.

A final photometric parameter fit is performed with the new
measurements, again with removal of data declared bad by the
cloud-camera veto, and with iterated outlier rejection.  The only
difference is that the time evolution $\dot{a}$ terms are allowed to
be non-zero.

Because we expect non-zero $\dot{a}$ terms to be caused by changing
atmospheric conditions, we experimented with allowing time evolution
in the extinction coefficients (eg, $\dot{k}$ terms).  Unfortunately,
such terms involve non-linear combinations of input data (time times
airmass) and can only be added without introducing biases if there are
*also* $\dot{a}$ terms; ie, it is wrong in principle to include
$\dot{k}$ terms without also adding $\dot{a}$ terms, especially in the
face of iterated outlier rejection.  We found that adding both
$\dot{a}$ and $\dot{k}$ terms did not improve our fits relative to
simply adding $\dot{a}$ terms, so we have not included $\dot{k}$
terms.  With more frequent standard-star sampling, the time resolution
of the system would be improved and $\dot{k}$ terms would, presumably,
improve the fits.

The entire next-day re-analysis takes between three and six hours,
depending on the amount of data.  Much of the computer time is spent
swapping processes in and out of virtual memory; with more efficient
code or larger RAM the re-analysis time could be reduced to under two
hours.  The re-analysis would take one to two hours longer if it were
necessary to repeat the astrometric solutions found in the real-time
analysis.

At the end of the re-analysis, the robot constructs a ``final'' WWW
page, similar to the real-time feedback WWW page, but with the final
photometric solution and residuals.  Parameters grossly out-of-spec
are shown in red.  Also, the robot sends an email to a SDSS email
archive and exploder, summarizing the final parameter values and rms
residuals in the five bands.

For the robot's purposes, a night is declared ``photometric'' if the
rms residual around the photometric solution (after iterated outlier
rejection at the 3-sigma level) in the $[u',g',r',i',z']$ bandpasses
is less than $[0.03,0.02,0.02,0.02,0.03]$~mag.  If the night is
declared photometric, then the bias, flat, and fringe images and their
associated pixel masks are declared ``current,'' to be used for the
real-time analysis of the following nights.

\begin{figure}
%
%
\plotone{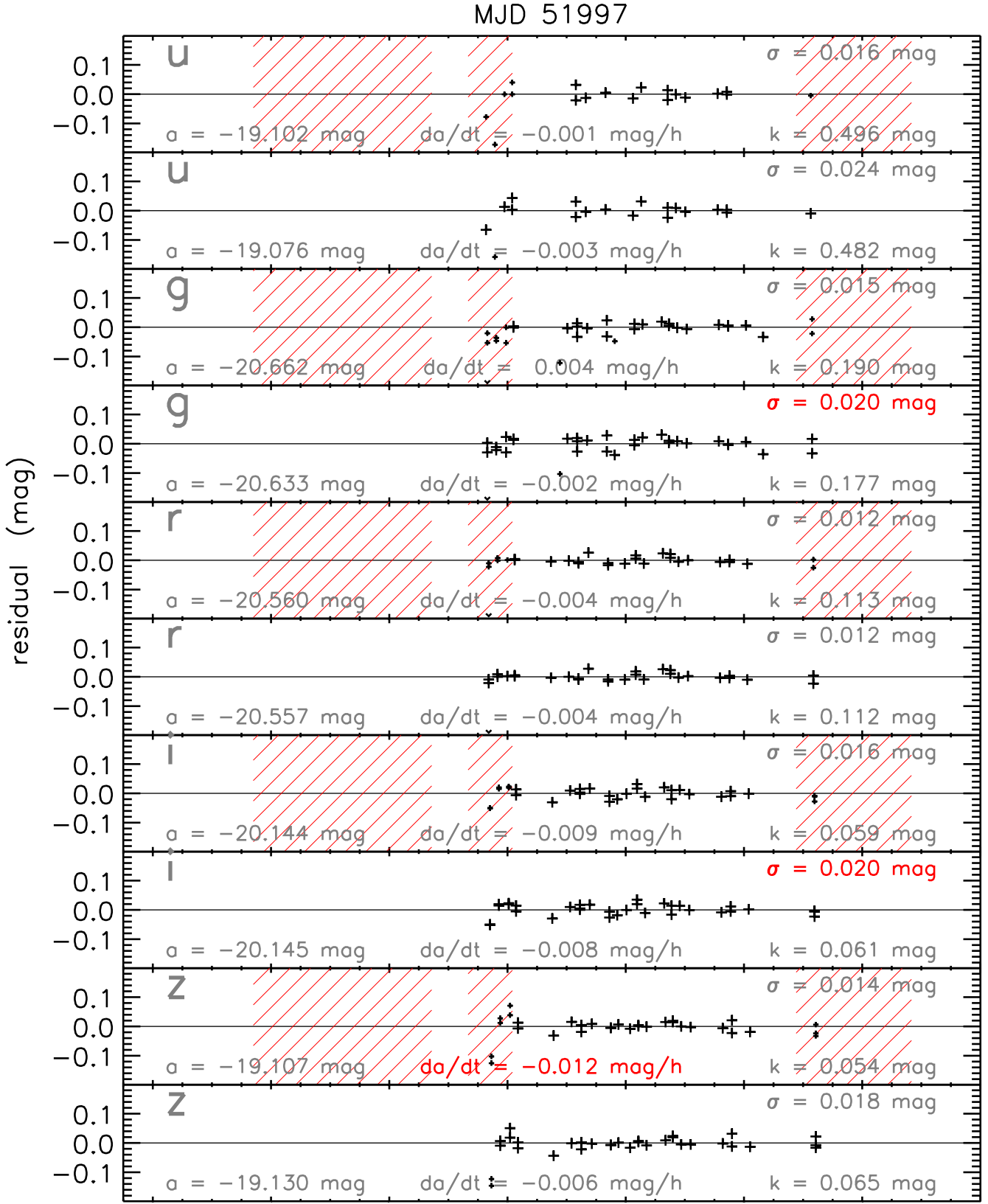}
\figcaption[Demonstration of cloud veto effectiveness]{A comparison of
final fits (ie, including $\dot{a}$ terms) in the five bands, for MJD
51997, with and without the data vetoing by cloud-camera.  The panels
with the red cross-hatching have had data taken during cloudy periods
removed from the fit.  In most bands, the fit is improved when cloudy
data are removed.\label{fig:cloudveto}}
\end{figure}

Figure~\ref{fig:cloudveto} shows the final fits for an example night,
with and without the inclusion of the veto of data taken during cloudy
periods.  This is a night which would have been declared marginally
non-photometric without the cloud-camera veto, but became photometric
when the cloudy periods were removed.

\section{Archiving}

All raw PT data are saved in a tape archive at Fermi National
Accelerator Laboratory (FNAL).  In addition, about one year's worth of
the most recent data are archived on the robot machine itself, on a
pair of large disk drives.  The photometric parameters from each
photometric night are kept in a FITS binary table (\cite{cotton95}),
labeled by observation date.  These parameter files are mirrored in
directories at FNAL and elsewhere (DWH's laptop, for example).
Because the astrometric solutions are somewhat time-consuming, the
astrometric headers for all the corrected frames are also saved on
disk on the robot machine and mirrored, along with each night's bias,
flat, and fringe images.  Because the astrometric header information is
all saved, along with the bias, flat, and fringe images, the corrected
frames can be reconstructed trivially and quickly from the raw data.

\section{Scientific output}

The archived output data from the photometricity monitor robot is not
just useful for verifying and analyzing contemporaneous data from the
Observatory.  It contains a wealth of scientific data useful for
analyzing long-term behavior and pathologies of the site, the
hardware, and the standard star catalog.  Many of these analyses will
be performed and presented elsewhere, as the site monitoring data
builds up.

\begin{figure}
%
%
\plotone{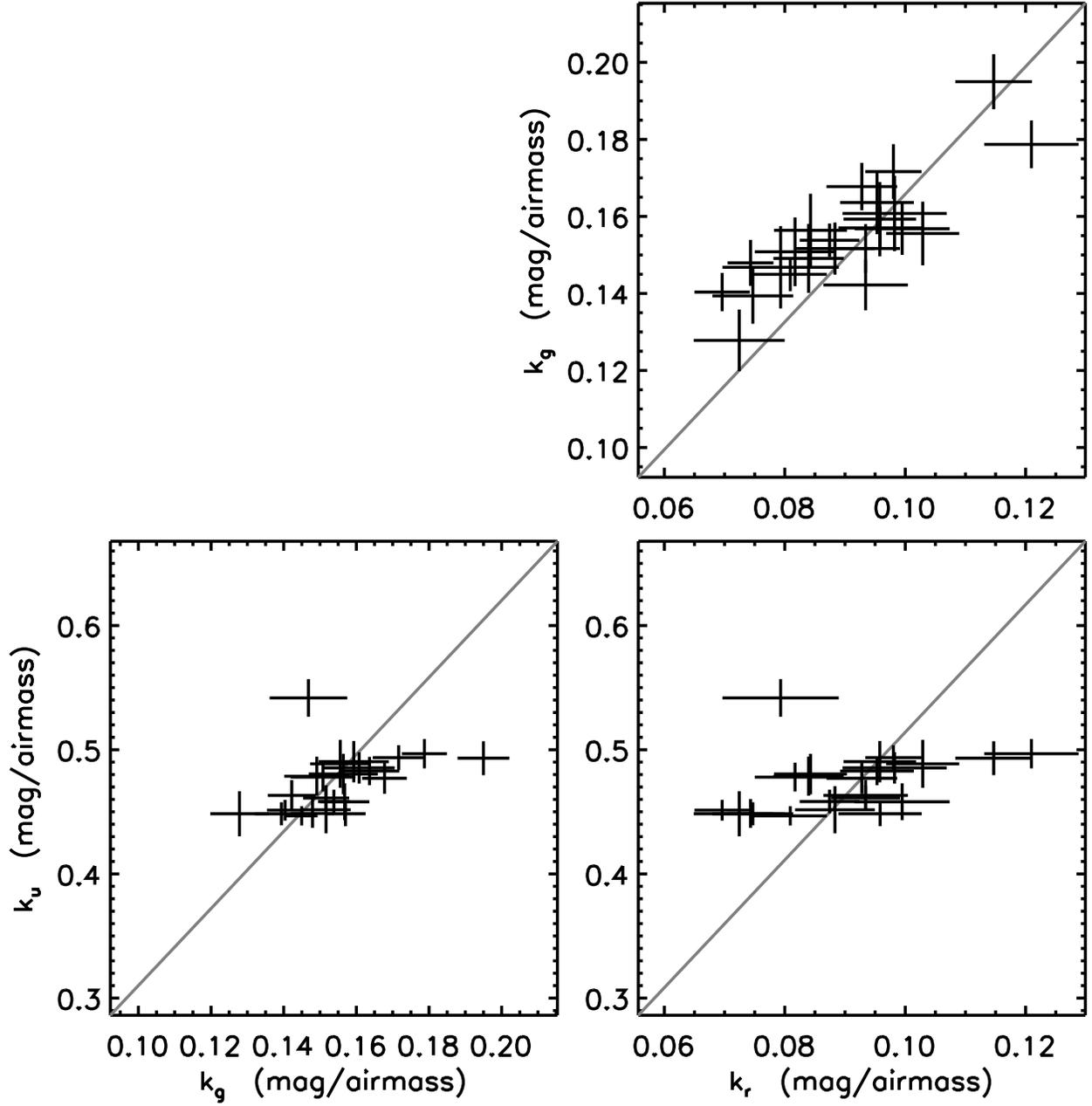}
\figcaption[atmospheric extinction covariances]{Atmospheric extinction
coefficients $[k_u,k_g,k_r]$ in the $[u',g',r']$ bandpasses, for most
photometric nights between MJD 51840 and 51970, plotted against one
another.  The grey lines show the relationship ($k_u\propto k_g\propto
k_r$) expected if atmospheric extinction is due to a single
atmospheric component.\label{fig:ext_covar}}
\end{figure}

As an example, Figure~\ref{fig:ext_covar} shows the atmospheric
extinction coefficients $[k_u,k_g,k_r]$ plotted against one another,
for most of the photometric nights in roughly one-third of a year.
This Figure shows the variability in the extinction coefficients, as
well as their covariance.  If the variability in the extinction is due
to varying optical depth of a single component of absorbing material,
the covariances should fall along the grey diagonal lines (ie,
$k_u\propto k_g\propto k_r$).  Although the $k_g$ vs $k_r$ plot is
consistent with this assumption, the $k_u$ vs $k_r$ plot appears not
to be.  Perhaps not surprisingly, atmospheric extinction must be
caused by multiple atmospheric components.  It is at least slightly
surprising to us that the variability in $u$-bandpass extinction is
smaller than would be predicted from a one-component assumption and
the variability in the $r$-bandpass extinction.

\section{Comments}

The success of this ongoing project shows that robust, hands-off,
real-time and next-day photometricity assessment and atmospheric
extinction measurement is possible.  There is much lore about
photometricity, site variability, and precise measurement in
astronomy.  Most of this lore can be given an empirical basis with a
simple system like the one described in this paper.

It is worth emphasizing that the observing hardware used by the robot
is \emph{not} dedicated to the robot's site monitoring tasks.  The PT
is being used to calibrate SDSS data; it only spends about one third
of its time taking the observations of catalogued standards which are
used for photometricity and extinction measurements.  This shows that
a robot of this type could, with straightforward (if not trivial)
adjustment, be made to ``piggyback'' on almost any observational
program, provided that some fraction of the data is multi-band imaging
of photometric standard stars.  The robot does not rely on the images
having accurate astrometric header information, or accurate text
descriptions or log entries; it finds standard stars in a robust,
hands-off manner.  Many observatories could install a robot of this
type with \emph{no hardware cost whatsoever!}  A site with no
appropriate imaging program on which the robot could ``piggyback''
could install a small telescope with a good robotic control system, a
CCD and filter-wheel, and the robot software described here, at fairly
low cost.  The telescope need only be large enough to obtain good
photometric measurements of some appropriate set of standard stars.
All of the costs associated with such a system are going down as
small, robotic telescopes are becoming more common (eg,
\cite{akerlof99}, \cite{strassmeier00}, \cite{querci00}).

The robot works adequately without input from the $10~\mathrm{\mu m}$
cloud camera, as long as data are not taken during cloudy periods, or
as long as the observers can mark, in some way accessible to the
robot, cloudy data.  On the other hand, $10~\mathrm{\mu m}$ pixel
array cameras, with no moving parts (unlike the prototype camera
working at APO) are now extremely inexpensive and would be easy to
install and use at any observatory.

The robot system was developed and implemented in a period of about
nine months.  It has been a very robust tool for the SDSS observers.
The rapid development and robust operation can be ascribed to a number
of factors: The robot design philosophy has always been to make every
aspect of the robot's operation as straightforward as possible.  We
have only added sophistication to the robot's behavior as it has been
demanded by the data.  The \texttt{IDL} data analysis language has
primitives useful for astronomy and it operates on a wide range of
platforms and operating systems.  Perhaps above all, the PT is a
stable, robust telescope (Uomoto et al in preparation).

Without objective, well-understood site monitoring like that provided
by this simple robot, analyses of archived and queue-observing data
will always be subject to some suspicion.  At APO, the visual monitor
robot has been a very inexpensive and effective tool for building
confidence in observer decisions and for providing feedback to data
analysis.  With this robot operating, APO has better monitoring of
site conditions and data quality than most existing or even planned
observatories.

\acknowledgements Comments, suggestions, data, computer code, bug
reports and hardware maintenance were provided generously by Bill
Boroski, Jon Brinkmann, Scott Burles, Bing Chen, Daniel Eisenstein,
Masataka Fukugita, Steve Kent, Jill Knapp, Wayne Landsman, Brian Lee,
Craig Loomis, Robert Lupton, Pete Newman, Eric Neilsen, Kurt
Ruthsmandorfer, Don Schneider, Steph Snedden, Chris Stoughton, Michael
Strauss, Douglas Tucker, Alan Uomoto, Brian Yanny, Don York, our
anonymous referee, and the entire staff of the Apache Point
Observatory.

The Sloan Digital Sky Survey (SDSS) is a joint project of The
University of Chicago, Fermilab, the Institute for Advanced Study, the
Japan Participation Group, The Johns Hopkins University, the
Max-Planck-Institute for Astronomy (MPIA), the Max-Planck-Institute
for Astrophysics (MPA), New Mexico State University, Princeton
University, the United States Naval Observatory, and the University of
Washington. Apache Point Observatory, site of the SDSS telescopes, is
operated by the Astrophysical Research Consortium (ARC).

Funding for the project has been provided by the Alfred P. Sloan
Foundation, the SDSS member institutions, the National Aeronautics and
Space Administration, the National Science Foundation, the
U.S. Department of Energy, the Japanese Monbukagakusho, and the Max
Planck Society. The SDSS Web site is http://www.sdss.org/.

Partial support for DPF has been provided by NASA.  This research made
use of the NASA Astrophysics Data System.

\begin{deluxetable}{rlccccc}
\tablenum{1}
\tablewidth{0pt}
\tablecaption{Approximate PT bandpass information\label{tab:bandpasses}}
\tablehead{
\colhead{}&\colhead{}&
\colhead{$u'$}&\colhead{$g'$}&\colhead{$r'$}&\colhead{$i'$}&\colhead{$z'$}
}
\startdata
central wavelength                 & (\AA) &
 3540      & 4770      & 6230      & 7620      & 9130 \nl
full-width at half-maximum         & (\AA) &
  570      & 1390      & 1370      & 1530      &  950 \nl
\enddata
\end{deluxetable}

\clearpage

\clearpage

\clearpage

\clearpage

\clearpage

\clearpage

\end{document}